\documentclass[a4paper,fleqn,usenatbib]{mnras}

\usepackage{newtxtext,newtxmath}

\usepackage[T1]{fontenc}
\usepackage{ae,aecompl}
\usepackage{gensymb}

\usepackage{graphicx}	
\usepackage{amsmath}	

\title[Anisotropic quenching in massive clusters]{Evidence for anisotropic quenching in massive galaxy clusters at $z\approx0.5$}

\author[J. P. Stott]{
John P. Stott,$^{1}$\thanks{E-mail: j.p.stott@lancaster.ac.uk (JPS)}
\\
$^{1}$Department of Physics, Lancaster University, Lancaster LA1 4YB, UK\\
}
\date{Accepted XXX. Received YYY; in original form ZZZ}

\pubyear{2019}

\begin{document}
\label{firstpage}
\pagerange{\pageref{firstpage}--\pageref{lastpage}}
\maketitle

\begin{abstract}
A recent observational result finds that the quenching of satellites in groups at $z=0.08$ has an angular dependence relative to the semi-major axis of the central galaxy. This observation is described as `anisotropic quenching' or `angular conformity'. In this paper I study the variation in the colour {of a mass limited sample of} satellite galaxies relative to {their angle from the} major axis of the Brightest Cluster Galaxy in the CLASH clusters up to $z\sim0.5$, 4\,Gyr further in lookback time. {The same result is found: }galaxies close to the major axis are more quenched than those along the minor axis. I also find that the star-forming galaxies tend to avoid a region $\pm45\degree$ from the major axis. This quenching signal is thought to be driven by AGN outflows along the minor axis, reducing the density of the intergalactic medium and thus the strength of ram pressure. Here I will discuss potential alternative mechanisms. Finally, I note that the advent of the Legacy Survey of Space and Time (LSST) and \emph{Euclid} surveys will allow for a more detailed study of this phenomenon and its evolution.

\end{abstract}




\section{Introduction}
\label{sec:intro}

Dense environments are known to quench the star formation within galaxies, which leads to the observation that a higher fraction of galaxies in massive clusters are passive early-types compared with the field (e.g. \citealt{dressler1980,peng2010}). The processes responsible for this must relate to the cluster environment, i.e. the massive dark matter halo, the hot gas of the intra-cluster medium (ICM) or the increased number density of the galaxy population.

Environmental quenching is thought to be due to a number of mechanisms. Quenching via ram pressure stripping occurs when a galaxy falls into the cluster and passes through the ICM. The force of the ICM gas on the cold gas in the disc of the galaxy removes this gas from the galaxy and thus the fuel for star formation \citep{gunn1972}. This may temporarily increase the star formation rate, causing a star burst \citep{dressler1983}, with star formation induced in both the gas compressed on the leading edge and that which is disturbed and pulled out into a tail behind \citep{fumagalli2014,poggianti2017}. Ram pressure stripping is thought to lead to rapid quenching on a timescale of $\sim1\,$Gyr regardless of whether the gas is removed and/or partially used up in a starburst \citep{roberts2019}. 

Instead of removing the cold gas from the disc the cluster environment may instead remove (or heat) the gas in the extended reservoir that surrounds the galaxy, stopping the replenishment of cold gas in the disc and therefore quenching the galaxy \citep{larson1980}, a phenomenon often known as `starvation' (e.g. \citealt{bekki2002}) or `strangulation' (e.g. \citealt{balogh2000}). This is a slower quenching process than ram pressure stripping, on a timescale of $\gtrsim3-4\,$Gyr \citep{roberts2019}, as the cold gas in the disc is unaffected and so can continue to form stars until it is exhausted. The original paper by \cite{larson1980} suggests that the gas surrounding galaxies will be removed by interactions with other galaxies in the cluster and their extended gas reservoirs. {However, the current view is that within the hot ICM, the galaxy's halo gas is kept hot or is ram pressure stripped by, or absorbed into, the ICM itself} and so is unable to cool and fall back down onto the disc where it can form stars \citep{bekki2002}. 

Galaxy major merging, which can cause quenching, is relatively rare in massive galaxy clusters due to the galaxies moving at speeds of $\gtrsim1000\,$km\,s$^{-1}$. However, gravitational interactions between galaxies can disturb the gas in spiral galaxies, initially creating a burst of star formation followed by subsequent quiescence \citep{moore1996}. Tidal effects due to the cluster potential as a whole may also produce a similar result (e.g. \citealt{byrd1990}).

Outflows and radiation from the supermassive black hole in the central galaxy can keep the ICM from cooling, thus suppressing the fuel for star formation on the central galaxy, an observation known as Active Galactic Nuclei (AGN) feedback (see \citealt{fabian2012} for a review). A further possibility is that the AGN activity of the central galaxy can act to suppress star formation in satellites. This has been seen in simulations, in which AGN feedback from the central galaxy increases the temperature and velocity of the intergalactic gas, therefore making it less suitable as a fuel for star formation in satellites \citep{dashyan2019}.

It has been observed that the star formation rate of the central galaxy and its satellites are linked, such that a quenched central galaxy will have quenched satellites and a star forming central galaxy will have star-forming satellites, a phenomenon termed `galactic conformity' \citep{weinmann2006}. In a recent paper, \cite{martnav2021} present the result that galaxy quenching in halos is anisotropic, such that the quenched galaxy fraction has an angular dependency; satellite galaxies that live along the minor axis of the central galaxy are less quenched than those along the major axis. This anisotropic quenching observation has been termed `angular conformity' by the authors of \cite{martnav2021}. The explanation they provide for anisotropic quenching, is that the AGN of the central galaxy preferentially acts along the minor axis creating bubbles in the halo gas and lowering the density. The lower density halo gas along the minor axis leads to a lowering of the strength of ram pressure acting on the satellites and therefore they are less quenched. This perhaps has the opposite effect of what one might naively expect for AGN activity acting along the minor axis. This is supported by their observation that the anisotropic quenching signal is stronger for satellites around centrals with the most massive black holes. They present further evidence for the AGN feedback connection, as the anisotropic quenching signal is also seen in the IlIustrisTNG simulations with improved AGN feedback treatment \citep{nelson2019}, while it is absent for the less realistic prescription employed in the original Illustris simulation \citep{nelson2015}. 

In their paper \cite{martnav2021} examine $\sim30$K groups and clusters containing a total of $\sim124K$ satellite galaxies, with a halo mass range of $\log$ (M/M$_{\odot}$) $= 12 - 14.5$ at a median redshift of $z=0.08$, from the Sloan Digital Sky Survey (SDSS, \citealt{sdss2014}). Here, I test whether it is visible in a much smaller sample of 13 rich clusters with halo masses $\log$ (M/M$_{\odot}$) $ > 14$,  observed with much deeper \emph{Hubble Space Telescope} (HST) photometry, enabling me to probe the anisotropic quenching signal in galaxies of $\log$ (M$_\star$/M$_{\odot}$)$>9.5$ out to $z\sim0.5$. The reason for testing this in massive clusters only, is because \cite{martnav2021} find a greater anisotropic quenching signal in the satellite population when in the presence of more massive and quiescent central galaxies. 

I adopt a cosmology with $\Omega_{\Lambda}$\,=\,0.7, $\Omega_{m}$\,=\,0.3, and H$_{0}$\,=\,70\,km\,s$^{-1}$\,Mpc$^{-1}$. All quoted magnitudes are on the AB system and I use a \cite{chabrier2003} IMF throughout. This research makes use of Astropy,\footnote{http://www.astropy.org} a community-developed core Python package for Astronomy \citep{astropy1, astropy2}.

\section{Sample and Data}
\label{sec:data}

The cluster sample was taken from the Cluster Lensing and Supernova Survey with Hubble (CLASH, \citealt{postman2012}) sample. The sample was initially reduced from the original 25 to 23 clusters by removing the two highest redshift clusters (MACSJ0744+39 at $z=0.686$ and CLJ1226+3332 and $z=0.890$), as they fall into a redshift range outside the main body of the sample and are therefore inappropriate for the consistent rest-frame photometry employed for the remaining clusters. 

The photometric catalogue was taken from \cite{molino2017}, which is available on the CLASH website. \footnote{https://archive.stsci.edu/prepds/clash/} More specifically, I used their {\sc photoz} photometry, which uses a consistent aperture across the wavebands and is therefore well suited to accurate colour determination. The photometry was corrected for galactic extinction using the \cite{schlegel1998} dust maps. I also used the \cite{molino2017} tabulated photometric redshifts and stellar masses of the galaxies, and the position angle and axis ratio of the Brightest Cluster Galaxy (BCG).

The central galaxy in the context of a galaxy cluster is the BCG. In order to compare with \cite{martnav2021} an unambiguous BCG with good photometric data was required. I therefore visually inspected all the clusters and their BCGs to further refine the samples, in order to identify the BCG and any problems associated with it. As a note of caution, there are columns for BCG coordinates in the published \cite{molino2017} catalogues but these appear to instead be the CLASH cluster centroid, which is \emph{either} the BCG coordinates \emph{or} the X-ray centroid and so it was important to perform this by-eye check. 

The visual inspection performed led to the further removal of several clusters by comparing the image with the \cite{molino2017} catalogues. The reasons for removing clusters were: the BCG contains dust lanes which significantly affect the photometry as it becomes two sources (Abell 383); the BCG contains multiple cores, which affect the photometry and the catalogue position angle (Abell 2261); the BCG is a merger which affects the photometry and potentially the catalogue position angle (MACSJ1931-26, MACSJ0329-02 and MACSJ1149+22); There are multiple BCGs, perhaps due to a cluster merger, and so the central galaxy is ambiguous (MACSJ0416-24, RXJ1347-1145, MACSJ0717+37 and MACSJ0647+70); The catalogue position angle appears incorrect (MACSJ0429-02); Strong lensing within the BCG affects the catalogue position angle (MACSJ2129-07). The outcome of this sample cleaning was that 13 CLASH clusters remained. 

{In order to determine whether a galaxy was quenched or not I used a colour composed of a filter either side of the the rest-frame 4000\AA\, break.} This colour correlates well with specific star formation rate (sSFR) for galaxies with a relatively low dust content, as it is essentially the UV flux divided by a redder optical flux (i.e. it is a star formation rate indictor divided by a stellar mass indicator). This allowed for comparison with \cite{martnav2021}, as they found that the anisotropic quenching signal was also seen in the average sSFR. The cleaned CLASH sample had a redshift range $z=0.206 - 0.545$ and so in order to {obtain consistent colours in the vicinity of the 4000\AA\, break}, I first split this into low and high redshift samples, which cover the ranges $z=0.206 - 0.352$ and $z=0.391 - 0.545$ and consist of 8 and 5 clusters respectively. The average masses of the samples, taken from \cite{merten2015}, are $8.6\pm2.0\times10^{14}$M$_\odot$ and $6.7\pm1.4\times10^{14}$M$_\odot$ for the low and high$-z$ samples respectively. {Finally, the filters used to straddle the 4000\AA\, break were} F390W-F625W and F475W-F814W respectively. {In \S\ref{sec:ana} I discuss the corrections applied to obtain a consistent colour for each cluster}. The final cleaned low and high$-z$ samples are provided in Table \ref{tab:sample}.

To determine cluster membership I designated any galaxy with a \cite{molino2017} photometric redshift within $\Delta z=0.05\times(1+z)$ of the cluster redshift to be a member, to account for the  uncertainty of the photometric redshifts. I also chose to include only galaxies with stellar mass $\log$ (M$_\star$/M$_{\odot}$)$>9.5$ as the mass is complete to approximately $\log$ (M$_\star$/M$_{\odot}$)$=9.0$ for the highest redshift cluster I use at $z=0.545$. Finally, for a fair comparison between clusters within the same redshift sample, a limiting radius from the BCG was used. This radius was set by the maximum radius in proper kpc, which fitted within the HST Wide Field Camera 3 (WFC3) field-of-view, for the lowest redshift galaxy cluster in the sample. This radial limit was 200kpc for the low$-z$ sample and 350kpc for the high$-z$ sample.


\begin{table*}
	
	\footnotesize
	\centering
	\caption{The final low and high redshift cluster samples.}
	\label{tab:sample}
	\begin{tabular}{lrrc} 
		\hline
		Cluster & BCG R.A. & BCG Dec. & $z$ \\
		 & (degrees) & (degrees) & \\
		\hline
low$-z$ sample &&&\\
Abell\,209			&	22.9690	&	-13.6112 & 0.206	\\					
Abell\,1423			&	179.3223	&	33.6110	& 0.213	\\	
RXJ\,2129+0005		&	322.4164	&	0.0892	& 0.234	\\	
Abell\,611			&	120.2367	&	36.0566	& 0.288	\\	 	
MS\,2137.3-2353	&	325.0632	&	-23.6611	& 0.313	\\	
RXJ\,1532.9+3021	&	233.2241	&	30.3498	& 0.345	\\				
RXJ\,2248-4431		&	342.1832	&	-44.5309	& 0.348		\\	
MACSJ\,1115+01	&	168.9663	&	1.4986	& 0.352		\\
\hline
high$-z$ sample &&&\\
MACSJ\,1720+35	&  260.0698	&	35.6073	& 0.391		\\		
MACSJ\,1206-08 	& 181.5507	&	-8.8009	&	0.440	\\
MACSJ\,0329-02	&  52.4232	&	-2.1962	&	0.450	\\
MACSJ\,1311-03		& 197.7575	&	-3.1777	&	0.494	\\
MACSJ\,1423+24	& 215.9495	&	24.0784	&	0.545	\\

		\hline
	\end{tabular}
\end{table*}

\section{Analysis and Results}
\label{sec:ana}

The main goal of this paper was to study the relationship between how quenched a satellite galaxy is and its angle on the sky compared with the major axis of the central galaxy. In practice I compared the colour of the cluster galaxies to their angle relative to the major axis of the BCG. As discussed in \S\ref{sec:data} I split the CLASH clusters into low and high redshift samples so that two different sets of observed filters could be used to measure {a consistent rest-frame colour in the vicinity of the 4000\AA\, break}. However, the clusters within the samples still had a range of redshifts ($z=0.206 - 0.352$ and $z=0.391 - 0.545$) {and so the probed rest-frame colour was affected by k correction and potentially galaxy evolution across these ranges. I therefore k and evolution corrected the photometry} to the central redshifts of the samples using a simple stellar population (SSP) model with a formation redshift of $z_f=2$, solar metallicity and a \cite{chabrier2003} IMF, from \cite{bruzual2003}. This model is appropriate to quenched cluster galaxies and not those with continued star formation, but as the redshift ranges were relatively small the corrections were suitable for the majority of the galaxies. 

\begin{figure*}
		\centering
		\includegraphics[width=\columnwidth, trim=0 0 0 0, clip=true]{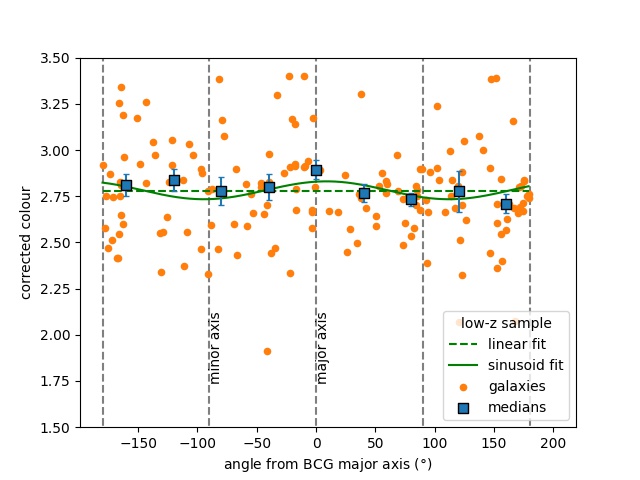}
		\includegraphics[width=\columnwidth, trim=0 0 0 0, clip=true]{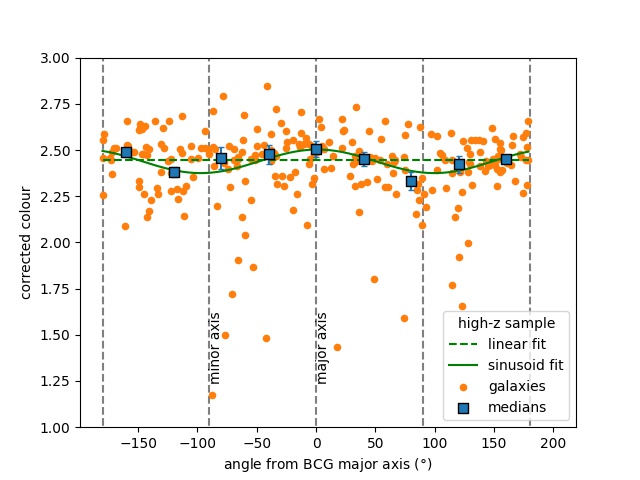}
    \caption[]{{\it Left}: {The corrected colour} plotted against the angle between the position of the satellite galaxies and the BCG major axis for the low redshift sample. The orange circle points are the individual galaxies, the blue squares are the median colour values in bins of 40 degrees. The solid green line is a sinusoidal fit to the median values and the dashed green line is a linear fit for comparison. {\it Right}: {The corrected colour} plotted against the angle between the position of the satellite galaxies and the BCG major axis for the high redshift sample. The symbols are defined in the same way as for the low redshift version. The high$-z$ sample is well described by the sinusoidal fit, whereas the signal is weaker for the low$-z$ sample.}
	    \label{fig:angq}
\end{figure*}

The angles between the satellite galaxies and the BCG major axis (provided in the \citealt{molino2017} catalogue) were then computed. {The colour is plotted against satellite galaxy} angle from the BCG major axis for both the low and high$-z$ samples in Fig. \ref{fig:angq}. A median colour value was calculated in bins of 40 degrees, with an error bar corresponding to the standard error. A sinusoidal least squares fit was then performed to these median values. For agreement with the observations of \cite{martnav2021} one would expect the fits to produce a cosine with a period of 180 degrees, with a peak coinciding with 0 degrees and a trough at 90 degrees, as I define 0 degrees to be along the major axis. The fit to the low$-z$ sample medians gives an amplitude of $0.048\pm0.029$\,mag, a period of $207\pm32$ degrees and the peak has a phase offset from the major axis of $15\pm30$ degrees. The amplitude of this signal is therefore only significant at the $1.7\sigma$ level. This fit has a reduced chi-squared value of $\chi_\nu^2=1.4$ which is the same value as for a horizontal straight line. This shows that the sinusoidal fit is not preferred to a straight line fit. The fit to the high$-z$ sample medians gives an amplitude of $0.064\pm0.020$\,mag, a period of $197\pm14$ degrees and the peak has a phase offset from the major axis of $2\pm12$ degrees. The amplitude of this signal is therefore significant to the $3.2\sigma$ level. This fit has a reduced chi-squared value of $\chi_\nu^2=0.9$, compared with $\chi_\nu^2=2.3$ for a horizontal straight line, which shows that the sinusoid is preferred. The difference between the cosine-fitted peak in colour compared with the trough is 0.13\,mag. This corresponds to a factor of $\sim1.5$ in sSFR, which I estimated empirically by fitting a straight line to the {relationship between the colour and sSFR for galaxies} with $\log$ (M$_\star$/M$_{\odot}$)$>9.5$ at $z=0.45-0.47$ (as the high$-z$ sample average $z=0.46$) from the Cosmic Assembly Near-infrared Deep Extragalactic Legacy Survey (CANDELS) catalogues of \cite{barro2019}. By examining Extended Data Figure 5 of \cite{martnav2021}, which shows sSFR plotted against satellite angle to the central galaxy major axis, I estimate the equivalent change in sSFR to be a factor of $\sim1.8$ and so the signal amplitude is comparable.

\begin{figure}
		\centering
		\includegraphics[width=\columnwidth, trim=0 0 0 0, clip=true]{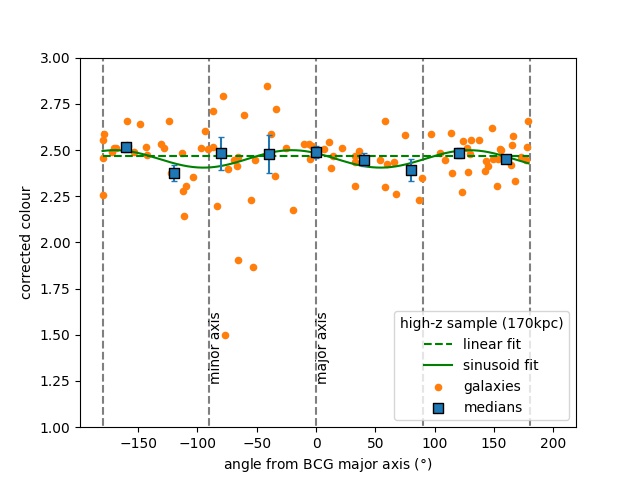}
    \caption[]{{The corrected colour} plotted against the angle between the position of the satellite galaxies and the BCG major axis for the high redshift sample but within a restricted analysis radius of 170\,kpc ($\sim$0.11\,R$_{200}$) to match the low$-z$ sample. The symbols are defined in the same way as for Fig. \ref{fig:angq}. In this case the signal is weaker, with a smaller period and a phase shift compared with the 350\,kpc radius version plotted in Fig. \ref{fig:angq}.}
	    \label{fig:angqsml}
\end{figure}

The above analysis shows that in the low$-z$ sample there may be a weak signal of anisotropic quenching but it appears stronger in the the high$-z$ sample. \cite{martnav2021} find that along the major axis, galaxies are quenched out to much larger cluster-centric radii than along the minor axis. Therefore, a possible reason for the discrepancy in the power of the signal between the low and high$-z$ samples is that the cluster-centric radius is 200\,kpc for the low$-z$ analysis and 350\,kpc for the high$-z$ sample due to the fixed angular size of the HST camera. This potentially means that the low$-z$ sample was not being probed to a large enough radius to see the reduction in quenching along the minor axis. The average R$_{200}$ of the low$-z$ and high$-z$ samples are 1.8\,Mpc and 1.5\,Mpc, and so on average I can only probe to 0.11\,R$_{200}$ and 0.23\,R$_{200}$ respectively.  I investigated the effect of radius by reducing the cluster-centric radius of the high$-z$ sample to 170\,kpc (i.e. 0.11\,R$_{200}$) for a fairer comparison. The result of this test is shown in Fig. \ref{fig:angqsml}, with the fit to the sample medians giving an amplitude of $0.047\pm0.019$\,mag, a period of $149\pm11$ degrees and the peak has a phase offset from the major axis of $49\pm29$ degrees. The amplitude of this signal is therefore only significant to the $2.5\sigma$ level and the period and phase offset are significantly different from the expected cosine wave. This suggests that the radius within which the quenching is quantified is important, with the radius needing to be large enough to observe the reduction in quenching along the minor axis in order to give a strong anisotropic signal. 

\begin{figure}
		\centering
		\includegraphics[width=\columnwidth, trim=0 0 0 0, clip=true]{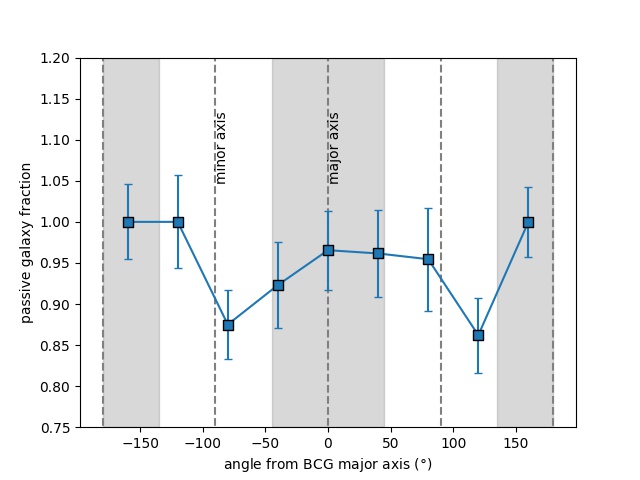}
    \caption[]{The ratio of the number of star-forming galaxies to passive galaxies plotted in $40\degree$ bins of angle to the BCG major axis for the high redshift sample. The grey shaded regions mark the $\pm45\degree$ either side of the major axis, which the star-forming galaxies appear to avoid.}
	    \label{fig:angrat}
\end{figure}

As well as being well fit by the sinusoid, one can also see that the bluer galaxies in the high$-z$ sample ({those below the red sequence with colours $C\lesssim2.0$}) appear to be less numerous in a range of angles within $\sim45$ degrees of the major axis, although this is not seen for the weaker signal of the low$-z$ sample. {This colour, $C=2.0$, corresponds to an sSFR of $\approx 2\times10^{-11}$yr$^{-1}$, based on an estimation from CANDELS as discussed above, which is an appropriate dividing line between passive and star-forming galaxies (e.g. Figure 2. of \citealt{davies2019})}. This is in agreement with the results presented in \cite{martnav2021}, who studied the fraction of quenched galaxies with angle instead of average colour, as I have presented in Fig. \ref{fig:angq}. In Fig. \ref{fig:angrat} I show the ratio of quenched galaxies to the total, as a function of angle from the BCG major axis for the high$-z$ sample, {assuming a colour of $C=2.0$} is the delimiter between star-forming and passive.

\section{Discussion}
\label{sec:disc}

As discussed in \S\ref{sec:intro}, \cite{martnav2021} explain that the observed anisotropic quenching signal is due to AGN outflows acting preferentially along the minor axis of the central galaxy, which lowers the density of the intergalactic medium in that direction, reducing the effect of ram pressure stripping on satellites, thus resulting in weaker quenching along the minor axis. This explanation is well supported by their hydrodynamic simulations. Their sample consists of 30K halos in the mass range M$_{200}=10^{12} - 10^{14.5}$\,M$_\odot$, and so is dominated by halos much less massive than the galaxy clusters I consider here. However, the analysis in \S\ref{sec:ana} demonstrates, at least for the high$-z$ sample, evidence of anisotropic quenching even in a relatively small sample of massive clusters. For this to be the case BCG radio jets that lead to bubbles which reduce the density of the ICM would need to preferentially occur along their minor axis. This is likely to be the case as radio galaxies classified as elliptical are found to have radio jets and lobes preferentially with $30\degree$ of their minor axis \citep{condon1991}.

The ability to detect the anisotropic quenching signal in such a small sample is likely because the CLASH clusters are some of the most massive known and therefore the environmental effects within them will be strong. This is particularly true for the combination of AGN feedback and ram pressure stripping given as a favoured mechanism by \cite{martnav2021}. The most massive clusters have the most massive central galaxies (e.g. \citealt{lauer2014}), with the most massive black holes (e.g. \citealt{merritt2001}), and the ICM is the most dense (e.g. \citealt{mohr1999}).

In \S\ref{sec:ana} I found a discrepancy between the low and high$-z$ samples as the anisotropic quenching signal is much weaker in the former. The difference between the mean redshifts ($z=0.287$ and $z=0.463$) corresponds to a difference in lookback time of 1.48\,Gyr, which is a relatively small timescale for such massive, mature clusters and so evolution cannot explain the discrepancy in signal strength. It appears from Fig. \ref{fig:angqsml} that this is instead due to the difference in cluster-centric radius between the samples (200 and 350\,kpc respectively), as \cite{martnav2021} find that the along the major axis, galaxies are preferentially quenched out to larger distances than along the minor axis and so there is little anisotropic quenching signal at smaller radii, where galaxies along both axes are quenched. While my study is limited by the size of the HST field, \cite{martnav2021} are uninhibited as they use SDSS. It is not clear what their maximum radius is and they perform the analysis in terms of virial radius in order to compare halos from M$_{200}=10^{12} - 10^{14.5}$\,M$_\odot$. They state that their satellite galaxies are at cluster-centric radii up to $\sim$900\,kpc and so assuming this is for their quoted most massive cluster with M$_{200}=10^{14.5}$\,M$_\odot$, this corresponds to $\sim$0.64\,R$_{200}$ and in other places in their paper 0.75\,R$_{200}$ is quoted. The largest radii I consider in this paper is 350\,kpc, which corresponds to $\sim$0.23\,R$_{200}$ and so one would expect a weaker signal than \cite{martnav2021} for that reason, as well as the significantly small sample size.

\begin{figure}
		\centering
		\includegraphics[width=\columnwidth, trim=0 100 0 0, clip=true]{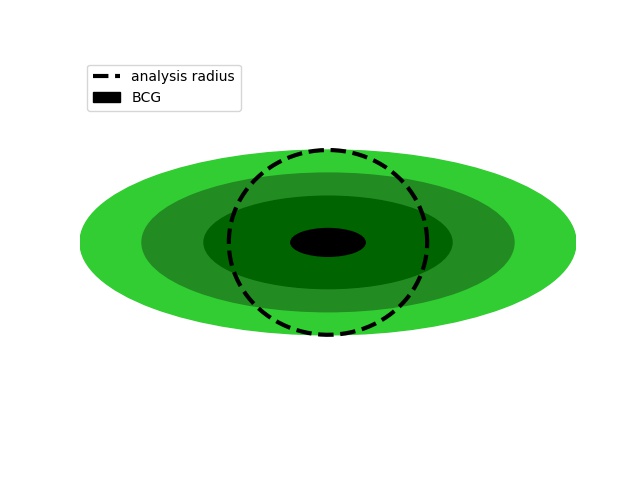}
    \caption[]{An illustration of the alternative explanations, other than AGN outflows, for the anisotropic quenching signal. In this example the BCG (black) and cluster (green) share the same significantly ellipsoidal shape and major axis. The green shades, dark to light, indicate decreasing density in either the ICM, dark matter or galaxy number density (or stellar mass density), which will lead to decreasing quenching of satellites. The black dashed circle is the cluster-centric radius within which the quenching signal is measured and so one would expect this to be weaker along the minor axis as compared with the major axis. }
	    \label{fig:diag}
\end{figure}

An alternative explanation for the anisotropic quenching could be the that the shape and major axis of the BCG tends to align with the shape and major axis of the cluster \citep{binggeli1982}. This major axis is thought to be along the major infall direction for matter and galaxies from the cosmic web (collimated infall) and this drives both the BCG and cluster to have the same alignment \citep{west1994}. In this scenario the ICM density of a significantly ellipsoidal or elongated cluster would be higher at a given cluster-centric radius along the major axis than the minor axis, even in the absence of an AGN. This would still reduce the magnitude of ram pressure stripping and strangulation/starvation, and therefore the quenching effect, and so when galaxy quenching is studied within a circular cluster-centric radius, this anisotropic quenching signal may appear. This would also be true of the dark matter density, which could lead to a reduction in quenching along the minor axis due to weaker tidal effects (e.g. \citealt{byrd1990}). Another related explanation is that the galaxies of the cluster align more along the BCG major axis because of the collimated infall direction, discussed above, and therefore the galaxy number density (or stellar mass density) is higher, leading to more gravitational interactions between satellites away from the minor axis \citep{moore1996}. I have included a basic illustration of the above scenarios in Fig. \ref{fig:diag}. I do not attempt to test these possibilities here as the CLASH data is too narrow-field but they are worth investigating despite the strong backing for the AGN outflow scenario seen in simulations. {A potential test of these scenarios would be to find clusters with apparent spherical halos from lensing but elliptical BCGs, to see if the result holds. However, it would be difficult to assess whether this was due to viewing the cluster at an angle which was straight down the infall axis. I note that the CLASH clusters are selected to be close to spherical, with an average ellipticity of $\epsilon=0.19$ \citep{postman2012,maughan2008} and so there is no evidence for strongly ellipsoidal or elongated clusters here.} The final possibility is also discussed by \cite{martnav2021}, which is that satellites along the minor axis may have somehow had their star formation enhanced by the AGN outflow from the central galaxy. However, there is little evidence for this in previous studies (e.g. \citealt{pace2014,lan2021})

\section{Summary \& conclusions}
\label{sec:conc}

In this paper the recently discovered phenomenon of anisotropic quenching (a.k.a angular conformity) was investigated using the CLASH cluster survey. This is the observation that satellite galaxies appear more quenched along the major axis of the central galaxy as opposed to the minor axis \citep{martnav2021}.

I looked for a periodic signal in the colour and quenched fraction {of a mass limited sample of satellite galaxies with} angle from the major axis of the BCG. There is evidence for such an anisotropic quenching signal in the high$-z$ sample at $z=0.391 - 0.545$ at the $>3\sigma$ level but only a weak signal in the low$-z$ sample at $z=0.206 - 0.352$. The explanation for this difference appears to be the smaller cluster-centric radius probed for the low$-z$ sample, imposed by the relatively narrow field of view of WFC3 camera. The amplitude of the anisotropic quenching signal is similar to that of \cite{martnav2021}, as I estimate a fit to their sSFR plot would give a factor of $\sim1.8$ change in sSFR, when comparing the satellites along the major axis to the minor axis. My fit indicates a factor of $\sim1.5$ change in sSFR.

Observing this signal out to $z\sim0.5$ pushes the study of this phenomenon 4\,Gyr further in lookback time than \cite{martnav2021}, who used the shallow but wide field area of SDSS to study 30K halos at $z\sim0.08$. This was made possible by the depth and resolution of the CLASH survey and the careful photometry and photometric redshifts of \cite{molino2017}.

\cite{martnav2021} have good evidence from their simulations that this signal is driven by AGN outflows along the minor axis, reducing the density of the intergalactic medium and therefore the strength of ram pressure on the satellites, leading to a directional reduction in quenching. I proposed some possible alternative explanations such as a reduction in ICM density and therefore ram pressure along the minor axis of the BCG as a natural consequence of the major axis and shape of the BCG and cluster tending to align. This alignment between the BCG and the cluster may also lead to a preferential quenching along the major axis due to tidal effects. Finally, due to the collimated infall and accretion onto the cluster from the cosmic web, the galaxy density along this major axis will be higher and therefore gravitational interactions between satellite galaxies, that lead to quenching, will be more likely.

Deep wide-field data with good quality photometric redshifts will be available in the coming years from the Legacy Survey of Space and Time (LSST) and \emph{Euclid} that will enable more detailed studies of this anisotropic quenching phenomenon, to discover the mechanism that drives it and potentially observe its evolution.

\section*{Acknowledgements}

{I would first like to thank the referee for their report, which improved the clarity of this paper. This research was prompted by attending a remote seminar given by Dylan Nelson to the Lancaster University, UK, Observational Astrophysics group on the 12th July 2021. This paper was entirely written on my dining room table due to the COVID-19 restrictions and guidelines in the UK. It is also my first paper using {\sc Python} for the analysis rather than {\sc IDL}. Finally, I would like to thank Harriet Rosenthal-Stott for proofreading this manuscript.}

\section*{Data availability}

The data underlying this article are publicly available from the CLASH cluster website (https://archive.stsci.edu/prepds/clash/) and the Mikulski Archive for Space Telescopes (MAST, https://archive.stsci.edu/hst/).




\bibliographystyle{mnras}
\bibliography{AngQuench} 




\appendix



\bsp	
\label{lastpage}
\end{document}